# The Global Impact Distribution of Near-Earth Objects

Clemens Rumpf [(1)], Hugh G. Lewis[(2)], and Peter M. Atkinson[(3)]

[(1)(2)] *University of Southampton, Faculty for Engineering and the Environment, SO171BJ, Southampton, UK*
[(3)] *Lancaster University, Faculty of Science and Technology, LA14YR, Lancaster, UK*
[(3)] *University of Utrecht, Faculty of Geosciences, Heidelberglaan 2, 3584CS, Utrecht, The Netherlands*
[(3)] *University of Southampton, Geography and Environment, SO171BJ, Southampton, UK*
[(3)] *Queen's University Belfast, School of Geography, Archaeology and Palaeoecology, BT71NN, Belfast, UK*

[(1)]C.Rumpf@soton.ac.uk
[(2)] H.G.Lewis@soton.ac.uk
[(3)]pma@lancaster.ac.uk

**Abstract**
Asteroids that could collide with the Earth are listed on the publicly available Near-Earth object (NEO) hazard web sites maintained by the National Aeronautics and Space Administration (NASA) and the European Space Agency (ESA). The impact probability distribution of 69 potentially threatening NEOs from these lists that produce 261 dynamically distinct impact instances, or Virtual Impactors (VIs), were calculated using the Asteroid Risk Mitigation and Optimization Research (ARMOR) tool in conjunction with OrbFit. ARMOR projected the impact probability of each VI onto the surface of the Earth as a spatial probability distribution. The projection considers orbit solution accuracy and the global impact probability. The method of ARMOR is introduced and the tool is validated against two asteroid-Earth collision cases with objects 2008 TC3 and 2014 AA. In the analysis, the natural distribution of impact corridors is contrasted against the impact probability distribution to evaluate the distributions' conformity with the uniform impact distribution assumption. The distribution of impact corridors is based on the NEO population and orbital mechanics. The analysis shows that the distribution of impact corridors matches the common assumption of uniform impact distribution and the result extends the evidence base for the uniform assumption from qualitative analysis of historic impact events into the future in a quantitative way. This finding is confirmed in a parallel analysis of impact points belonging to a synthetic population of 10006 VIs. Taking into account the impact probabilities introduced significant variation into the results and the impact probability distribution, consequently, deviates markedly from uniformity. The concept of impact probabilities is a product of the asteroid observation and orbit determination technique and, thus, represents a man-made component that is largely disconnected from natural processes. It is important to consider impact probabilities because such information represents the best estimate of where an impact might occur.

# 1 Introduction
An asteroid impacting the Earth is typically not amongst the concerns of people in everyday life. Nonetheless, the asteroid threat is real (Brown et al. 2002) and can have disastrous consequences. Asteroids have hit the Earth since the formation of the solar system and this process continues today. The bolide over Chelyabinsk in February 2013 that injured more than 1500 people demonstrated this palpably (Popova et al. 2013). The scientific community and leading nations broadly recognize that the asteroid hazard is a significant threat to our civilization. A result of this recognition is the establishment of international organizations (UN Office for Outer Space Affairs 2013) to address the threat and commencement of the search for potentially Earth-colliding objects (National Research

Council et al. 2010). The products of the search for asteroids are publicly available Near-Earth object (NEO) webpages, which list potentially impacting asteroids, and are maintained by the European Space Agency (ESA)[1] and the National Aeronautics and Space Administration (NASA)[2]. These lists include all known asteroids that have a notable chance of impacting the Earth in the next century but the impact distributions of these asteroids on the Earth's surface are not published.

Previous research has addressed the topic of the impactor distribution on the surface of the Earth, but none performed a quantitative assessment of the distribution. Three datasets that relate to the impact distribution problem and that are based on natural asteroids are known to the authors. They are based on historical impact records. The Russian "Institute of Computational Mathematics and Mathematical Geophysics" maintains the "Expert Database on Earth Impact Structures" (EDEIS) containing over 1000 impact crater features (EDEIS et al. 2006) and the results are available as a map that shows the locations of these impact features (confirmed and possible). However, the geological traces of impacts on Earth suffer from erosion and many might have disappeared or are not easily detected. In addition, water impacts rarely leave long-lived traces. Consequently, the data is biased towards land impacts and also depends on the localized interest of the population or scientists to identify impact features, both of which introduce additional bias into the database. Therefore, EDEIS data are not suitable for a global impact distribution analysis nor has such a quantitative assessment been undertaken. A similar dataset with the same limitations is maintained by the Planetary and Space Science Centre of the University of New Brunswick in Canada (University of New Brunswick 2014). The third dataset is comprised of airburst recordings obtained by a global infrasound microphone network. The network's original objective was to monitor atmospheric nuclear weapons tests, but since the signature of an airbursting asteroid shares enough similarities with that of a nuclear test, the network is able to detect these natural events and triangulate the airburst location. NASA has published these data in the form of a map that is based on recordings in the 1994-2013 timeframe (NASA 2014). These data are limited to the size regime of asteroids that experience an airburst when colliding with the Earth, but coverage is global and no bias is expected in the detection method. However, no quantitative assessment of the impact distribution based on these data has been performed; the reason might be that the data are too sparse to support such assessment.

An alternative to using observed asteroids as the basis for impact distribution analysis is to use a representative, artificial population of virtual impactors and such a synthetic population was generated by Chesley & Spahr (2004). The work focuses on the impactor distribution in a celestial, geocentric coordinate system that uses spherical coordinates with the ecliptic as reference and the Earth-Sun opposition point as origin. Notably, the work finds that most impactors approach roughly from the orbit or anti-orbit direction of the Earth with a minor concentration approaching from the opposition direction. Furthermore, it was shown that the majority of impactors are concentrated in the ecliptic plane. In contrast to the research presented here, the celestial frame does not rotate with the Earth and, thus, the locations on the Earth that the impactors would impact are not apparent, nor was this the aim of the study. In Grav et al. (2011) an impact location map of North America was shown that is based on the synthetic population, but without quantitative assessment of the impact location distribution.

In the research presented here, the future, potential impact distributions of observed asteroids were calculated and analyzed. The key motivating question was whether the impact distribution is uniform or if some regions are more likely to be hit than others. In Chesley & Spahr (2004), it was shown that impactors are expected to approach the Earth from directions that are roughly parallel to the Earth's orbit and that the majority of potential impactors reside in the ecliptic plane. In conjunction with the observation that the Earth performs a daily rotation under this constant influx, it can be asserted that all longitudinal sections of Earth are equally exposed to impacts and that the impact distribution in the longitudinal direction is uniform. However, an intuitive understanding of the latitudinal distribution of impact locations is not as easily obtained. Considering that most impactors are expected to originate from close to the ecliptic plane their impact velocity vectors should also be approximately parallel to the ecliptic plane. Further, assuming that the impactor influx density in the ecliptic normal direction is constant over the width of the Earth, the highest impact location density would be expected near the equator because the Earth surface bends away from the impactor influx towards the poles. This concept is depicted in Figure 1. Of course, impactors do not impact the Earth on a straight line: rather, the gravitational attraction of Earth bends the impactor's trajectory towards Earth. Consequently, impactors that would miss the Earth in the absence of Earth's gravitational field actually impact because their trajectory is changed under the influence of Earth's gravity. This means that the uneven distribution on the Earth expected without gravity is attenuated somewhat towards a more even distribution because gravitationally captured impactors from outside the physical diameter of the Earth impact closer to the poles (than the equator) resulting in a more balanced near-polar impact density (Figure 1).

---

[1] ESA NEO Coordination Centre webpage: http://neo.ssa.esa.int/
[2] NASA Near Earth Object Program: http://neo.jpl.nasa.gov/risks/

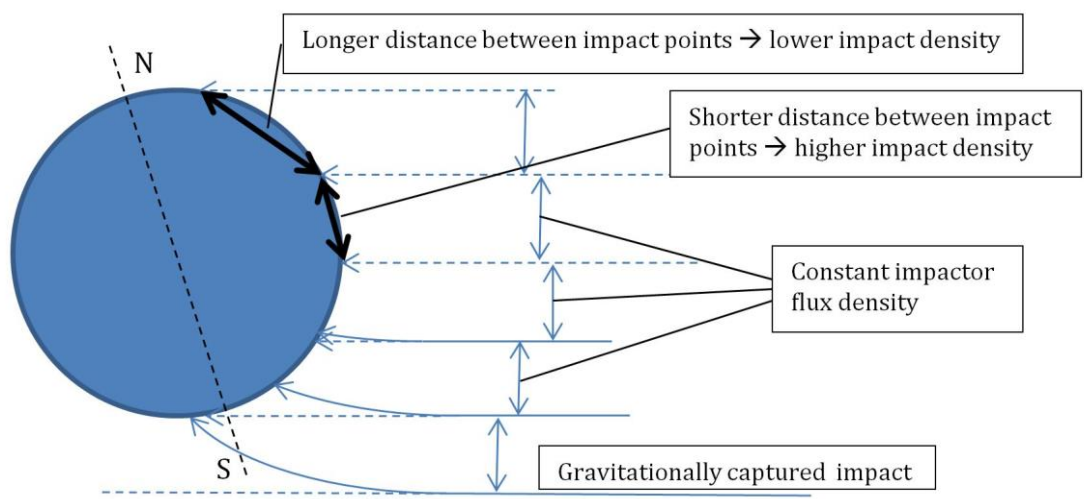

*Figure 1: Depiction of intuitive impactor density distribution in latitudinal direction neglecting the attracting effect of the gravity of Earth (dashed impactor lines) and under gravitational attraction ("Gravitationally captured impact", solid lines).*

To assess if the impact location distribution is uniform, the impact locations of 261 potential impactors (belonging to 69 observed asteroids), which can collide with the Earth before the year 2100, were calculated in a dynamic solar system simulation and visualized. The considered asteroids had a diameter range of an estimated[3] 30 m to 341 m. For comparison, the Chelyabinsk event was associated with a 19 m sized asteroid (Popova et al. 2013; Borovicka et al. 2013) while the devastating 1908 Tunguska event was likely caused by a 30 m sized object (Boslough & Crawford 2008; Chyba et al. 1993).

**2 Method**
The nominal orbital solution of an asteroid is a state vector describing the asteroid's orbit and position that fit best the observations that are available for this asteroid. A covariance matrix represents the uncertainty region that is associated with the orbital solution. The uncertainty region has a weak direction, commonly referred to as Line of Variation (LOV), along which the asteroid position is only poorly constrained and it typically stretches along the orbit of the asteroid (Milani et al. 2005). Using the data of available observations and the current nominal orbital solution of an asteroid that are provided on the ESA NEO webpage, the freely available software OrbFit (Milani et al. 1997) was utilized to identify orbit solutions that lie on the LOV as well as inside the uncertainty region and that result in a future Earth impact. The 69 asteroids were sampled from the ESA NEO webpage at random in the October 2014 timeframe. OrbFit samples the uncertainty region to find these impacting orbit solutions that are called virtual impactors (VI). It should be noted that one asteroid may have multiple impact possibilities in the future and thus yields more than one VI.

---
[3] Asteroid sizes are estimates based on their brightness and the values were taken from the ESA risk webpage.

The Asteroid Risk Mitigation Optimization and Research (ARMOR) tool was used subsequently to project the impact probability of these VIs onto the surface of the Earth. ARMOR used the VI orbit solution from OrbFit as the initial condition for the trajectory propagation until impact. Each VI propagation was started 10 days before impact and utilized a solar system model that considered gravitational forces from the Sun, the barycenters of the planetary systems and Pluto as well as point sources for the Earth and the Moon. The positions of the attracting bodies were retrieved from a lookup table that is based on the JPL DE430 planetary ephemerides (Folkner et al. 2014) and the interpolation scheme achieves millimeter level accuracy. The resulting gravitational acceleration $\ddot{r}$ for the VI, calculated in an inertial frame with origin in the solar system barycenter, is given by:

$$\ddot{r}_j = \sum_{i=1}^{11} -\frac{GM_i}{|r_{ij}|^3} r_{ij} \qquad Eq\ 1$$

where subscript $j$ denotes the VI, subscript $i$ denotes the attracting body, $r_{ij}$ is the position vector connecting the attracting body to the VI and $GM_i$ is the gravitational constant of the attracting body. The gravitational differential equation is numerically solved using the variable time step, predictor-corrector ADAMS method of the Livermore Solver for Ordinary Differential Equations (LSODE) (Radhakrishnan & Hindmarsh 1993) package that is used in the Python *scipy.integrate.odeint* function. Figure 2 shows the position discrepancy of the propagator compared to NASA's HORIZONS system (Giorgini & Yeomans 2013) for three asteroids over a 50 day period. In the application presented here, the propagator was used for propagation times of 10 days and the position error remains well bounded to within 50 m in this time frame.

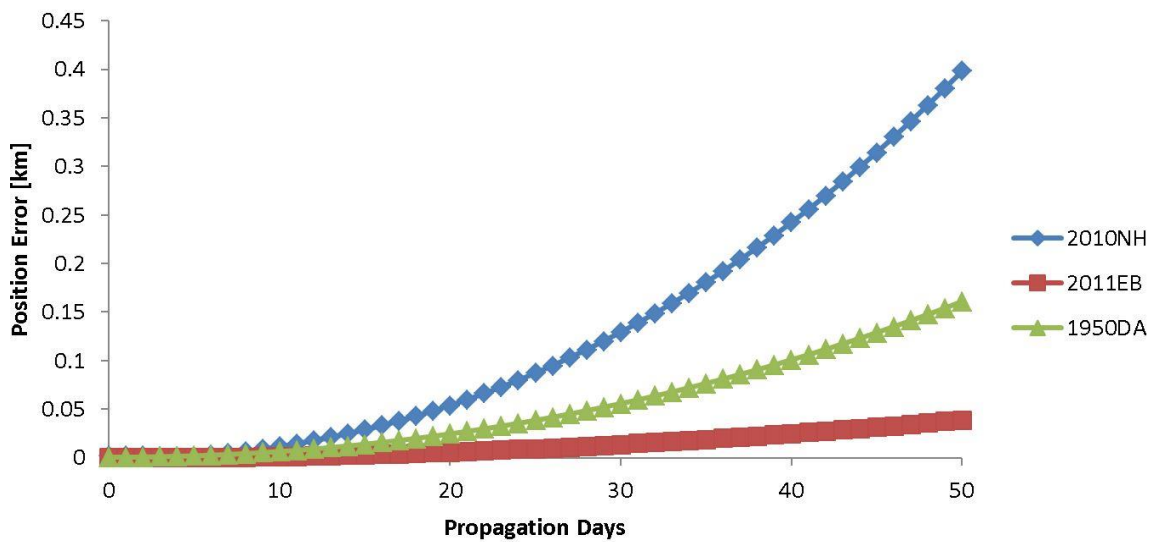

*Figure 2: Propagator accuracy comparison with JPL's HORIZONS system for 3 asteroids over a 50 day period. Note that the error does not grow exponentially but shows linear growth with sinusoidal characteristics for longer propagation times.*

ARMOR's propagator provides the trajectory of the VI and the center of the Earth as a set of $x, y, z$ coordinate time histories. The approximate impact time and location are determined by comparing the coordinates of the Earth and the VI. In the vicinity of the impact, the trajectories of the Earth and the VI are expressed by three-dimensional polynomials with time as the independent variable, which facilitates determination of impact points in analytical form. The method uses least squares to fit each coordinate component to a third order polynomial of the form:

$$P = c_0 + c_1 t + c_2 t^2 + c_3 t^3 \qquad Eq\ 2$$

For Earth's $x$ component, the expression is:

$$P_{\{♁,x\}} = c_{\{0,♁,x\}} + c_{\{1,♁,x\}} t + c_{\{2,♁,x\}} t^2 + c_{\{3,♁,x\}} t^3 \qquad Eq\ 3$$

Where $P_{\{♁,x\}}$ is the polynomial expression for the Earth's (♁) $x$ component, $c_{\{0,♁,x\}}$ is one of the polynomial coefficients and $t$ is time as the independent variable. Similarly, the asteroid's (✻) $x$ component is:

$$P_{\{✻,x\}} = c_{\{0,✻,x\}} + c_{\{1,✻,x\}} t + c_{\{2,✻,x\}} t^2 + c_{\{3,✻,x\}} t^3 \qquad Eq\ 4$$

The other position polynomials, $P_{\{♁,y\}}$, $P_{\{♁,z\}}$, $P_{\{✻,y\}}$ and $P_{\{✻,z\}}$ are determined accordingly. The next step is to subtract the Earth and asteroid polynomials corresponding to each Cartesian component. This is demonstrated here for the $x$ component:

$$P_{\{♁✻,x\}} = P_{\{♁,x\}} - P_{\{✻,x\}} \qquad Eq\ 5$$

$$P_{\{♁✻,x\}} = (c_{\{0,♁,x\}} - c_{\{0,✻,x\}}) + (c_{\{1,♁,x\}} - c_{\{1,✻,x\}}) t + (c_{\{2,♁,x\}} - c_{\{2,✻,x\}}) t^2 + (c_{\{3,♁,x\}} - c_{\{3,✻,x\}}) t^3 \qquad Eq\ 6$$

In accordance with Pythagoras, the difference polynomials for each component were squared and summed to produce the square of the distance, $D_{\{♁✻\}}$, between the Earth and the asteroid:

$$D^2_{\{♁✻\}} = P^2_{\{♁✻,x\}} + P^2_{\{♁✻,y\}} + P^2_{\{♁✻,z\}} \qquad Eq\ 7$$

The result of this computation is $D^2_{\{♁✻\}}$ and is a polynomial in itself. To find the time of impact of the asteroid, the real roots of Eq 7 are determined after the distance $D_{\{♁✻\}}$ is set equal to Earth's radius (6371 km) plus 42 km to model the surface of the atmosphere. In general, more than one real root can be found. One root is the time when the asteroid penetrates the sphere (6413 km radius) of the Earth, this is the impact time, and another root is the time when the asteroid exits the sphere of the Earth. Additional roots may be found that depend on the polynomial behavior outside the time interval of interest in the vicinity of the impact time. To determine the correct real root, the time of closest approach is considered. The time of closest approach is found by differentiating Eq 7 with respect to time. The first derivative of Eq 7 evaluates to zero at the time of closest approach. The correct time of impact is the real root closest to the time of closest approach that is also smaller than the time of closest approach. The time of impact serves as input to the positional polynomials of the asteroid ($P_{\{✻,x\}}, P_{\{✻,y\}}$ and $P_{\{✻,z\}}$) to obtain the precise impact coordinates. The obtained impact time determines the sidereal hour angle of the Earth (Curtis 2012) which, together with the impact coordinates, allows calculation of the impact's latitude and longitude. This is the impact point of the VI solution.

Orbit solutions that are located in the immediate vicinity of the VI and on its LOV also impact the Earth. To construct the impact corridor, the LOV is sampled by varying the epoch associated with the VI's orbital solution. The sampled orbit solutions produce impact points that form the impact corridor together with the impact point of the VI. Sampling by varying the VI's orbit solution epoch is equivalent to assuming that the LOV stretches in a similar direction as the VI's velocity vector. In the "vast majority of cases" this is a reasonable assumption (Milani et al. 2005) and the calculated impact corridor will match the real impact corridor.

To validate the impact corridor calculation of the ARMOR tool, three case studies were selected representing asteroids 2011 AG5, 2008 TC3 and 2014 AA. In the case of 2011 AG5 successful cross-validation with other predictive software tools was accomplished. Additional details on this case are provided in (Rumpf 2014).

Asteroid 2008 TC3 was discovered shortly before entering the Earth's atmosphere. Its entry point was predicted and the resulting bolide was observed by eye witnesses, satellite and infra-sound sensors (Jenniskens et al. 2009; Chesley et al. 2014). For validation, ARMOR used the nominal orbital solution for 2008 TC3, as fitted by OrbFit based on the available pre-impact asteroid observations, and predicted the atmospheric entry point as well as the ground track (Figure 3). The predicted nominal entry point agreed to within 0.39° longitude and 0.12° latitude (corresponding to a positional discrepancy of 44.3 km) at 65.4 km altitude. Furthermore, the shape of the ground track agreed well with the literature.

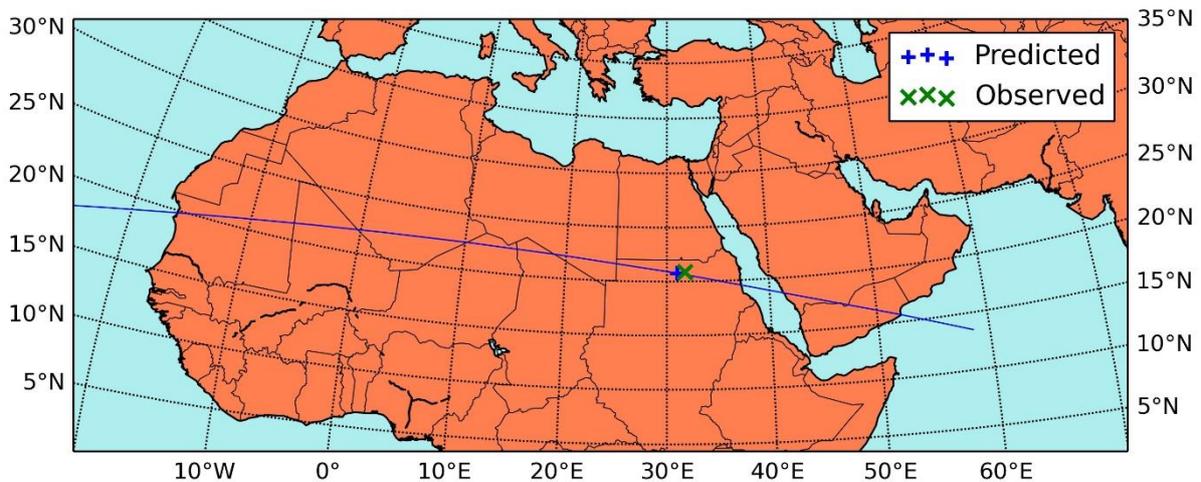

*Figure 3: Observed and predicted entry point for asteroid 2008 TC3. The blue plus sign marks the point of the nominal entry solution while the green cross gives the solution of the observed entry point at the same altitude of 65.4 km.*

Similarly to 2008 TC3, asteroid 2014 AA was discovered a few hours before its collision with the Earth. Based on the available observations, the entry point could be constrained to lie in the southern Atlantic Ocean. With the help of a global network of infra-sound microphones that recorded and triangulated the impact event, the most likely impact location was determined at the coordinates 44.207° longitude west and 13.118° latitude north (Chesley 2015). ARMOR predicted the entry point corresponding to the nominal, best fit orbital solution, provided by HORIZONS, at 46.42° longitude west and 12.98° latitude north. This corresponds to a discrepancy of about 240.2 km (with most of the deviation along the line of variation) (Figure 4).

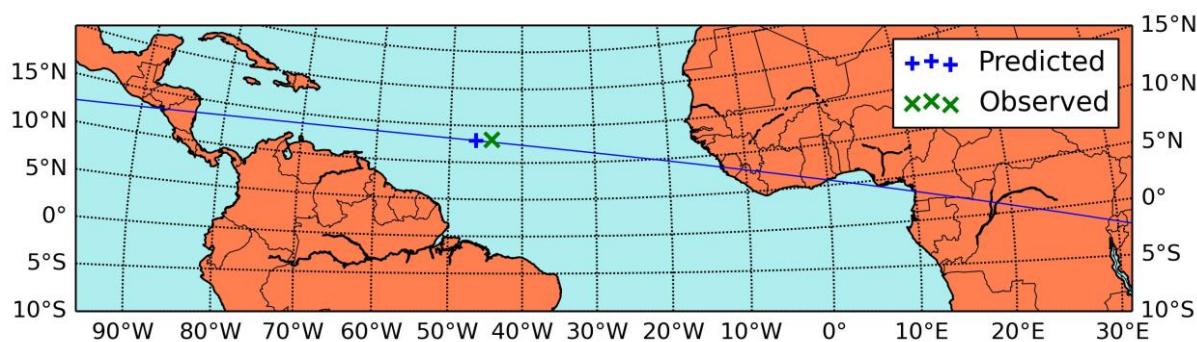

*Figure 4: Observed and predicted entry point for asteroid 2014 AA. The blue plus sign marks the point of the nominal entry solution while the green cross gives the solution of the observed entry point at the same altitude of 50 km.*

The positional discrepancy between observed and predicted impact points can be explained by several factors. The asteroid ephemeris is not known perfectly and the real and propagated trajectories differ through this error in the initial conditions. The propagator has an inherent propagation error that results in a deviation from the real trajectory. ARMOR uses a spherical Earth model (radius = 6371 km) and does not presently account for the oblateness of the Earth. This means that the impact location is effectively calculated at a different altitude than the observed one. Given a non-perpendicular impact trajectory with respect to the local horizon, the difference in altitude will produce a position error in the horizontal plane. Furthermore, ARMOR does not account for atmospheric interaction in the trajectory calculation while a real asteroid will experience drag and lift forces during atmospheric entry that affect the flight path of the asteroid. It is also expected that thermo-chemical interactions such as ablation and the resulting mass loss affect the asteroid trajectory and these effects are not modelled in ARMOR. Given the modelling constraints of ARMOR, the predicted impact points are reasonably close to the observed ones and the validation cases demonstrate that the impact point and corridor line calculations produce plausible results.

The impact corridor line forms the center line of the impact probability corridor projected on the Earth. A normal distribution with a 1-sigma value equal to the LOV width (a parameter available on the NEO webpages) is centered on the impact corridor line to represent the cross track impact probabilities. This newly formed impact probability distribution is scaled so that its integral is equal to the impact probability of the VI.

## 3 Results and Discussion

The method was applied to all 261 potential VIs and the result is a set of impact corridors, each in the form of a Gaussian distribution, that reflects the impact probabilities of the assessed VIs. All impact solutions were combined within a global map and the result is shown in Figure 5.

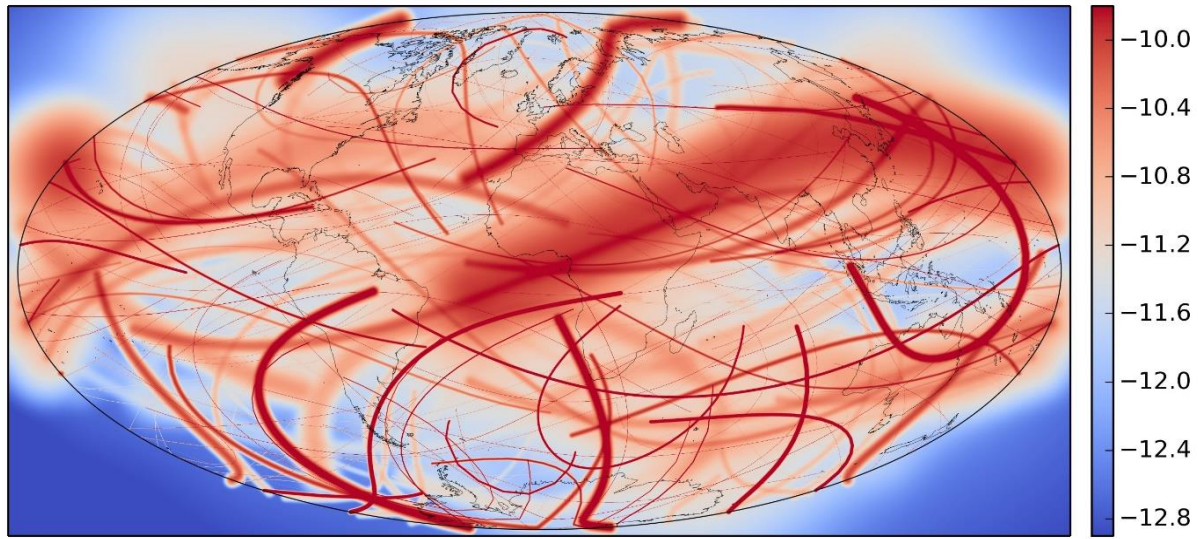

*Figure 5: The Earth in the Hammer projection showing the impact probability distributions for 261 VIs. The colour coding represents the impact probability at each location using a logarithmic scale (shown right).*

It is noticeable that the impact probability distributions in Figure 5 cover the entire Earth and that an asteroid impact can potentially happen anywhere. The literature indicates that the impact distribution of asteroids is uniform (NASA 2014) based on historical impact records. However, results in Figure 5 clearly show that some areas experience higher impact probability than others. To consolidate both remarks, one needs to recognize the difference between impact probability distribution and impact corridor distribution.

The impact corridor distribution shows where it is physically possible for the asteroids to impact, while the probability distribution also provides information about where an impact is more likely based on observational data. To visualize the impact corridor distribution, each virtual impactor was assigned an impact probability of one as well as a corridor width of 0.01 Earth radii and this representation of the impact corridor distribution is shown in Figure 6. Through this representation, it becomes apparent that impact corridors are more evenly distributed than suggested in the impact probability distribution shown in Figure 5. A quantitative assessment of the impact distribution is presented in the following.

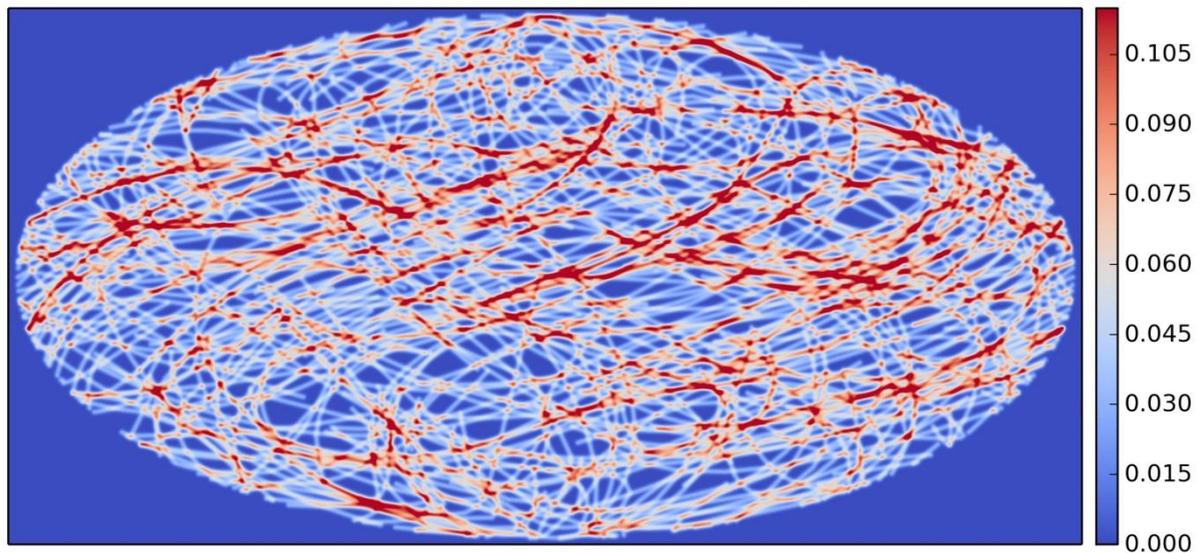

*Figure 6: Representation of all impact corridors showing their uniform global distribution. The impact probability of each VI was set to one and the corridor width was set to 0.01 Earth radii to facilitate equal visualisation of all impact corridors.*

Three probability density distributions are shown in Figure 7 as a function of longitude (top plot) and latitude (bottom plot). Uniform impact probability, as suggested by the literature, is represented by the blue line. Impact probability distribution, sampled in Figure 5 is represented by the red line. The green line shows the impact corridor distribution from Figure 6. Data sampling was accomplished using a fixed latitude (longitude) step width of 0.0357° to satisfy the Nyquist-Shannon limit. The maps used in this paper have a constant pixel area corresponding to 21.16 km$^2$ and the data may, thus, be compared between different regions and across maps as each pixel represent the probability density at the sampled coordinate. Finally, all three distributions have been normalized such that the integral of each distribution is one. This allowed them to be placed on the same scale for easier comparison.

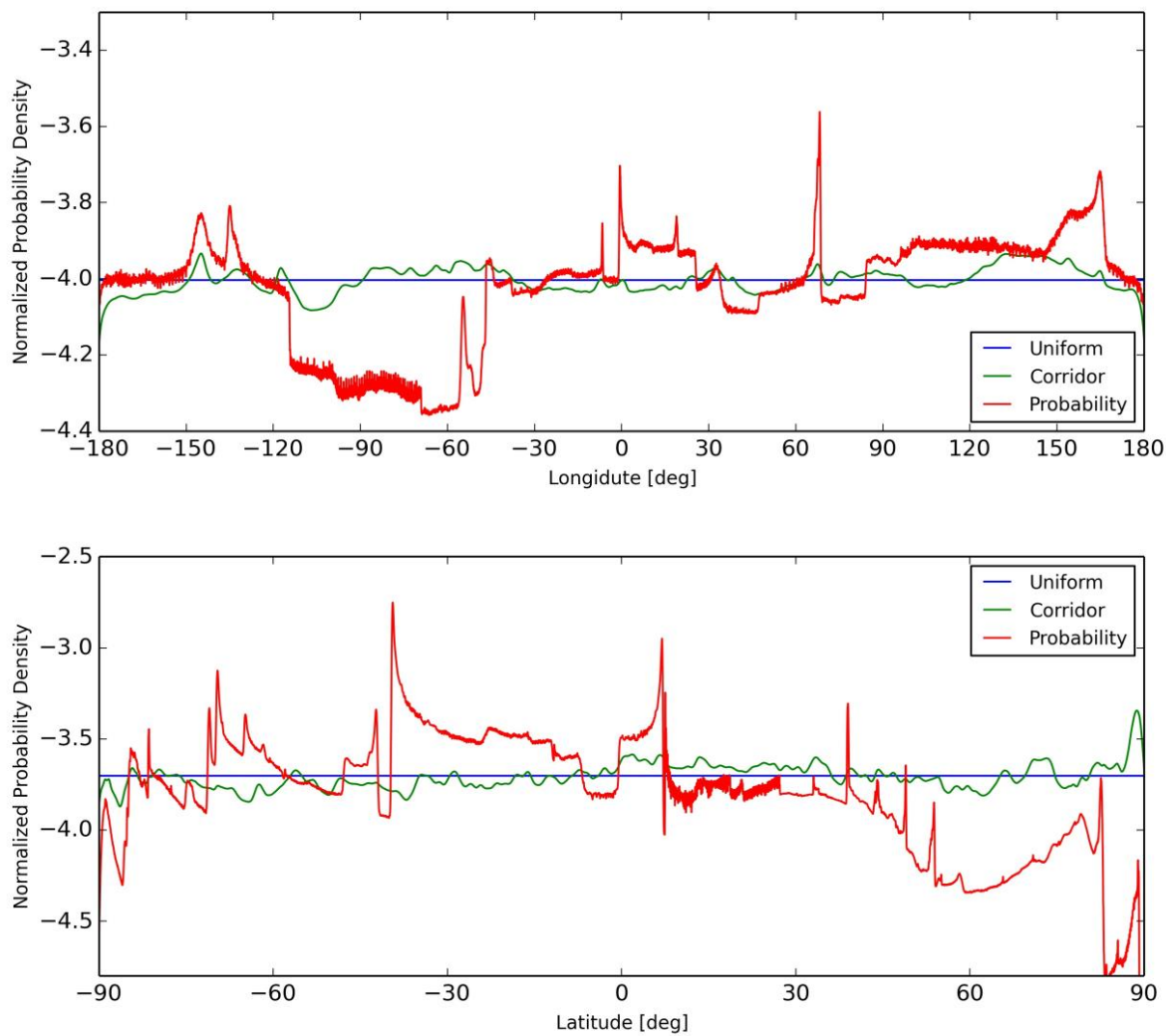

*Figure 7: Normalized probability density distribution over (upper plot) longitude and (lower plot) latitude. The blue line shows the expected result for a uniform distribution. The green line shows the calculated distribution of impact corridors assuming equal impact probability for all corridors. The red line indicates the distribution of corridors accounting for their predicted impact probabilities. Data are shown on a base 10 logarithmic scale.*

Although the normalization of the plots in Figure 7 lessens some of the information carried by the numbers, it facilitates easy comparison between the distributions by evaluating differences between them. The predicted distribution of impact corridors (green line) agrees closely with the assumption of uniform impact probability (blue line) with a root mean square difference (RMS) of $7.627 \times 10^{-6}$ for longitude and $3.452 \times 10^{-5}$ for latitude. On the other hand, the impact probability distribution differs more considerably from uniformity and the RMS increases by an order of magnitude to $2.834 \times 10^{-5}$ for longitude and $1.427 \times 10^{-4}$ for latitude.

To enable a comparison between the two independent datasets, the synthetic population of 10006 virtual impactors, mentioned in the introduction of this paper (Chesley & Spahr 2004), was analysed for its impact distribution. To generate the discrete impact distribution based on the synthetic population, the VI impact points were plotted on the world map. In accordance with the method used on ARMOR's set, this map was subsequently sampled for the spatial impact density distribution and the result has been normalized and plotted on the same scale as ARMOR data in Figure 7. Figure 8 shows three distributions in two plots for longitudinal (top) and latitudinal (bottom) direction. In these plots, data shown in grey represents the discrete impact distribution and the dashed, blue line is the uniform distribution (same as in Figure 7). A moving average filter with a 2° window width was used to smooth the discrete impact distribution and the result is represented by the red line. The RMS difference between discrete impacts (grey) and the uniform distribution (blue) is $7.229 \times 10^{-5}$ in the longitudinal direction and $2.096 \times 10^{-4}$ in the latitudinal direction. This result may be compared to the corridor distribution of the ARMOR sample and one notices that the ARMOR sample has an order of magnitude smaller RMS difference. The key difference between the synthetic VI set and the ARMOR VI set is that the former consists of 10006 discrete impact locations while ARMOR treats each of its 261 VIs as a probability distribution, effectively providing thousands of sample points per VI, and ARMOR's corridor distribution, thus, appears smoother (grey in Figure 8 vs green in Figure 7). In contrast to the discrete impact distribution in the longitudinal direction, which appears homogeneous over the entire longitudinal range, the distribution in latitude shows a recognizably different pattern near the poles (-90°/+90°) and it seems as if the impactor density increases towards the poles. However, analyses using the moving average filter reveals that the impact distribution remains, in fact, constant and that the peculiar appearance of the distinct impact distribution near the poles is due to increased variability in the data. The reasons for this increase in variability are that the area that corresponds to each latitude section ring decreases towards the poles and fewer VIs impact in these areas. Both aspects yield some empty section rings and some to accrue higher densities, effectively allowing for small sample artefacts to occur near the poles and this reflects in the bottom plot of Figure 8.

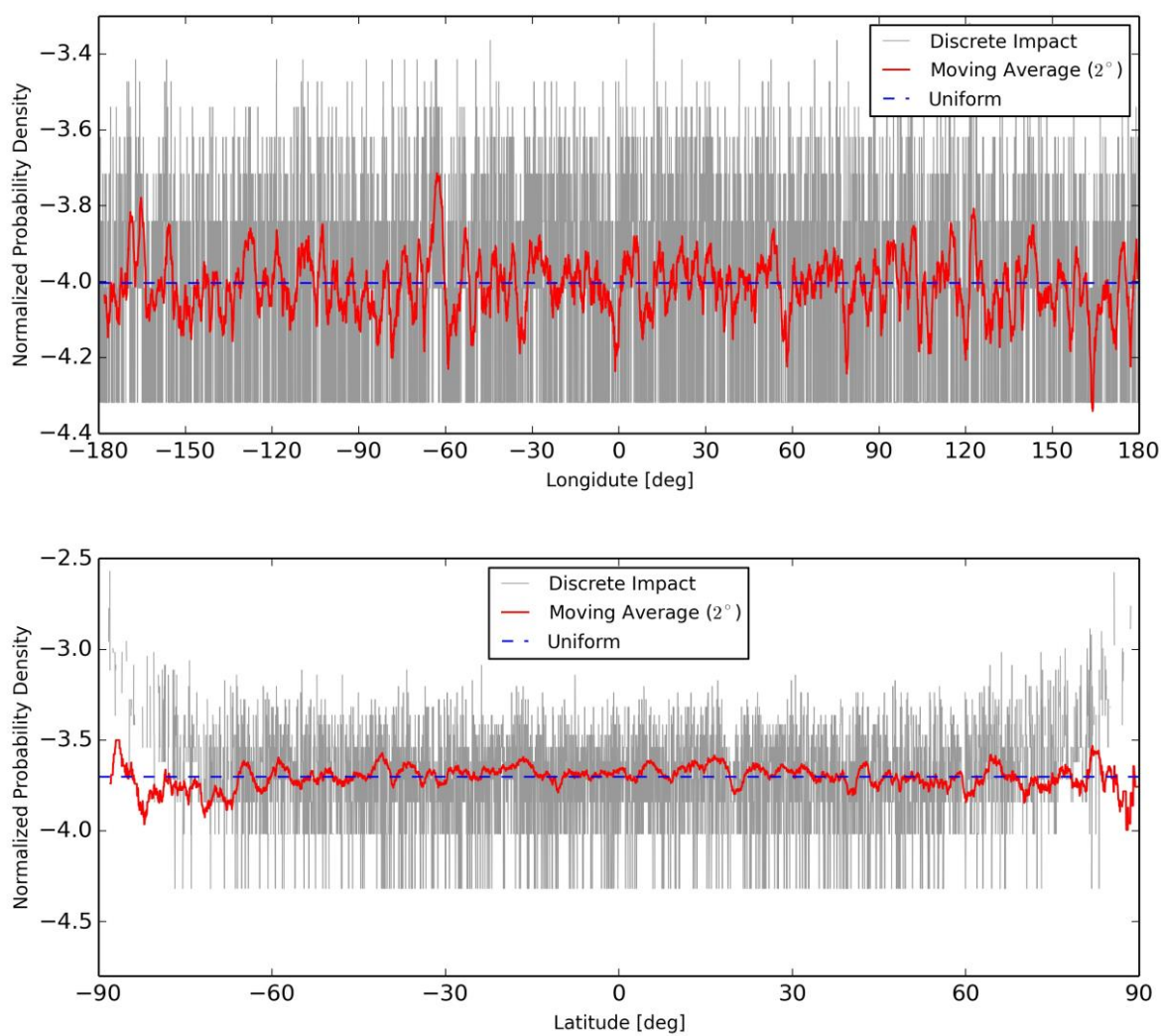

*Figure 8: Impact distribution based on a synthetic virtual impactor set comprising 10006 impact points. Grey represents the discrete impact distribution, red is the result of a moving average filter with 2° window width that was used on the discrete impact data and blue is the uniform impact distribution. Data is shown on a base 10 logarithmic scale.*

The two results that are based on the ARMOR sample and on the synthetic VI sample agree with each other and both results reinforce the assumption that the natural impact distribution is uniform. Furthermore, the results show that the sample size used for the analysis in ARMOR (Figure 7) is sufficiently large for the analyzed VIs to approach the natural, uniform impact distribution (blue and green line). This was expected intuitively for the distribution in the longitudinal direction. However, the results could not as easily be predicted for the impact density distribution in the latitudinal direction. The finding appears to confirm that the mechanics that lower the impact distribution density towards the poles (due to impactors originating from close to the ecliptic plane; Figure 1) are cancelled by the gravitational attraction of impactors by Earth.

Natural impact distribution is determined by the occurrence of asteroids in the Solar System and their trajectories. However, the VIs that produce the impact corridors are not only analyzed for their potential impact locations but also for their currently predicted impact probability. The predicted impact probability depends on the measurement technique that entails number, timing and quality of observations that are available for the parent asteroid as well as the orbit determination procedure. Thus, in addition to the natural population of NEOs and their trajectories, the results in Figure 5 depend on technology and scientific techniques that are disconnected from the natural processes that produce the impact distribution. The measurement and orbit determination process introduces a large amount of variability in the results because the predicted impact probabilities vary by orders of magnitude and this is why the impact probability distribution in Figure 5 (and red line in Figure 7) does not reflect a uniform distribution (blue line in Figure 7). The fact that the probability distribution is not uniform indicates limitations in the data.

It is estimated that only about 1% of all NEOs in the relevant size regime have been discovered (Harris & D'Abramo 2015). The impact probability distribution in Figure 5 reflects the current best knowledge of 69 analyzed asteroids which represent about 15% of all known asteroids with a non-zero impact probability (519 at the time of this writing[4]). Consequently, Figure 5 represents only a subset of the asteroids that may potentially impact the Earth in the future. It is possible that a larger sample size will approach uniform impact probability distribution. However, the variability in the probability distribution covers about five orders of magnitude (maximum impact probability: $7.4 \times 10^{-4}$; minimum impact probability: $9.69 \times 10^{-9}$). It is not clear if a large enough sample will ever be available to smoothen out the variability in impact probabilities. This is because new observations may not only add potentially impacting asteroids to the sample when new objects are discovered but also remove them by confirming that a potential impact is, in fact, impossible. Thus, the sample size does not necessarily grow over time with the availability of new observations. Furthermore, the discovery of a new high impact probability asteroid would increase the variability in the distribution (e.g. Asteroid 99942 "Apophis" has had a 2.7% impact probability in the past (Giorgini et al. 2008)) and skew the impact probability distribution towards its impact corridor. The impact probability distribution is time dependent and may never reach uniform distribution. The result in Figure 5 is, thus, a snapshot of the current situation based on our knowledge, and it will change over time. The result is representative of the type of information, with respect to impact probability distribution, that will be available in the future and has relevance as such: it represents humankind's present best guess at where an impact might occur.

**4 Conclusions**
This paper presented the impact corridor and probability distributions of 261 VIs belonging to 69 asteroids that currently have a chance of colliding with the Earth. Furthermore, it analysed the distributions for their conformity with the uniform impact distribution assumption. The distributions are calculated and visualized using the ARMOR software tool that can project impact probabilities of known asteroids onto the surface of the Earth. ARMOR's method was outlined and validation cases for its propagation accuracy as well as impact point and corridor calculation were presented. In this respect, the validation cases demonstrated that the tool produces plausible results. In addition, an analytical approach was developed to find the asteroid's impact point and time. Results show that the natural impact corridor distribution supports the assumption of uniform impact distribution and, crucially, extends the evidence basis for this assumption from past impact records into the future. This finding is confirmed by parallel analysis of impact points belonging to a synthetic population of 10006 VIs. However, in contrast and relative, to the corridor distribution, the deviation of the impact probability distribution from uniformity increased by an order of magnitude because of the large variation in VI-specific impact probabilities. The concept of impact probabilities is a product of the asteroid observation and orbit determination technique that is more dependent on technology rather than natural processes. Therefore, it introduces man-made variability of several orders of magnitude into the distribution. It is important to recognize this difference to set expectations about the characteristics of impact probability distributions that may be produced in the future. It can be concluded that, every region on Earth has a similar, a priori likelihood of an asteroid impact but that humankind's knowledge about the current asteroid impact situation is limited. Nevertheless, it is worth considering probabilistic data for practical purposes because, in the event of a recognized asteroid threat, a response decision will, in part, be based on imperfect knowledge such as that shown.

---
[4] ESA NEO Coordination Centre webpage accessed 05 August 2015: http://neo.ssa.esa.int/

**Acknowledgement and Funding**
The authors would like to extend their sincere thanks to Giovanni B. Valsecchi for the productive discussions and assistance in the use of OrbFit. Peter M. Atkinson is grateful to the University of Utrecht for supporting him with The Belle van Zuylen Chair. The work is supported by the Marie Curie Initial Training Network Stardust, FP7-PEOPLE-2012-ITN, Grant Agreement 317185. The authors would like to thank Dr Steven Chesley who provided the synthetic VI population.